\documentclass[floatfix,aps,prb,twocolumn,showpacs,groupedaddress]{revtex4}
\usepackage{graphicx}
\usepackage[]{epsfig}
\usepackage{times}
\newcommand{\nil}{\hspace*{0em}}
\newcommand{\beq}{\begin{equation}}
\newcommand{\eeq}{\end{equation}}
\newcommand{\bea}{\begin{eqnarray}}
\newcommand{\eea}{\end{eqnarray}}
\newcommand{\beas}{\begin{eqnarray*}}
\newcommand{\eeas}{\end{eqnarray*}}
\newcommand{\nn}{\nonumber}

\newcommand{\eps}{\epsilon}
\newcommand{\veps}{\varepsilon}
\newcommand{\al}{\alpha}
\newcommand{\s}{\sigma}
\newcommand{\lam}{\lambda}

\newcommand{\om}{\omega}

\newcommand{\imag}{{\rm Im}}

\begin{document}
\bibliographystyle{apsrev}
\title{Flux-dependent Kondo temperature\\
in an Aharonov-Bohm interferometer with an in-line quantum dot}

\author{Pascal Simon$^{1,4}$}
\author{O. Entin-Wohlman$^{2,3,4,}$ }\altaffiliation[On
leave from  ]{ the School of Physics and Astronomy, Raymond and
Beverly Sackler Faculty of Exact Sciences, Tel Aviv University,
Tel Aviv 69978, Israel.}
\author{A. Aharony$^{2,4,}$  }\altaffiliation[On leave from  ]
{ the School of Physics and Astronomy, Raymond and Beverly Sackler
Faculty of Exact Sciences, Tel Aviv University, Tel Aviv 69978,
Israel.}

\affiliation{$^{1}$ Laboratoire de Physique et Mod\'elisation des
Milieux Condens\'es, CNRS et Universit\'e Joseph Fourier, 38042
Grenoble, France  \\} \affiliation{$^2$ Department of Physics, Ben
Gurion University, Beer Sheva 84105, Israel\\} \affiliation{$^3$
Albert Einstein Minerva Center for Theoretical Physics  at the
Weizmann Institute of Science, Rehovot 76100, Israel\\}
\affiliation{$^4$ Materials Science Division, Argonne National
Laboratory, Argonne, Illinois 60439, USA}

\date{\today}
\begin{abstract}

An Aharonov-Bohm interferometer (ABI) carrying a quantum dot on
one of its arms is analyzed. It is found that the Kondo
temperature of the device depends strongly on the magnetic flux
penetrating the ring. As a result, mesoscopic finite-size effects
appear when the Kondo temperature of the dot on the ABI is
significantly smaller than the nominal one of the quantum dot
(when not on the interferometer), leading to plateaus in the
finite-temperature conductance as function of the flux. The
possibility to deduce the transmission phase shift of the quantum
dot from measurements of the ABI conductance when it is opened
(i.e., is connected to more than two leads) is examined, 
leading to the conclusion  that finite-size effects, when
significant, may hinder the detection of the Kondo phase shift.

\end{abstract}


\maketitle

\section{Introduction}
The inherent wave nature of the electron can be  revealed in
mesoscopic solid-state interferometers. These are built with
narrow wave guides for the electronic paths to preserve the
coherence of the electron. In general, the Aharonov-Bohm
interferometer (ABI) consists of a ring, which is threaded by a
Aharonov-Bohm (AB) magnetic flux $\Phi$.  When the ABI is
connected to two reservoirs, via two leads, it is termed `closed',
since  all the current  entering it through one lead leaves it
through the other. The `open' ABI connects to additional leads
(through which electrons can go astray, so that current is not
conserved). The conductance of the interferometer oscillates with
the flux, due to interference of the electronic wave function
between the two branches of the ring.\cite{imry} These AB
oscillations of the conductance have been first observed on
metallic two-terminal rings (i.e.,  closed
interferometers).\cite{webb85} Another stringent manifestation of
the electronic wave interference is the prediction of a
circulating (persistent) current around the interferometer even
when the latter is not connected to any
reservoir.\cite{buttiker85} These theoretical predictions have
been tested experimentally by various groups in the last two
decades.\cite{pc}

Recent experiments have used the ABI, when connected to electronic
reservoirs, as a tool to probe quantum coherent transport at the
mesoscopic scale. In these experiments, quantum dots are embedded
either on one
arm\cite{yacoby95,schuster,sprinzak,kouwenhoven,ji,kobayashi} or
on both arms\cite{holleitner,fuhrer} of the interferometer. These
experiments have triggered a large series of theoretical analyses
(see Ref. \onlinecite{hackenbroich} for a review). Transport
through a quantum dot can be characterized by a complex
transmission amplitude $t_{QD}=|t_{QD}|e^{i\varphi_{QD}}$. The AB
interferometry  promises a way to measure information on the
phase of the quantum dot, for example by measuring the
conductance oscillations through an open ABI which obeys 
certain conditions.\cite{schiller,abh} This information is
particularly interesting when the quantum dot is tuned in the
Kondo regime, where it mimics the behavior of an artificial
magnetic impurity.\cite{kastner,ji,cronenwett,blick,kouwenhoven}
At sufficiently low temperature, this artificial spin-1/2 impurity
is screened by the conduction electrons to form  a singlet, and
then the scattering of the conduction electrons on the impurity is
predicted to suffer a $\pi/2$ phase shift, associated with the
so-called unitary limit.\cite{nozieres} A series of recent
experiments on open multi-channel solid-state ABI's  \cite{ji}
have attempted to deduce this Kondo phase shift from the
dependence of the conductance oscillations on the dot gate voltage
and other parameters. These experimental results were not in
agreement with earlier theoretical predictions based on the exact
solution of the Anderson model, \cite{costi} and consequently
several attempts have been made to reconcile theory with the
experiment, some questioning the universality of the measured
phase. \cite{schiller,abh,aharony04} The theoretical understanding
of the experiments in Ref. \onlinecite{ji} is still under
debate.\cite{lavagna}

Nonetheless, all the recent theoretical analysis of the transport
through an ABI containing one or two quantum dots in the Kondo
regime \cite{costi,bulka,hofstetter,konig,ora04,lopez05,entin04}
ignores the finite-size extension of the so-called Kondo cloud.
Finite-size effects occur for example when the (artificial)
magnetic impurity is embedded into some finite-size
box.\cite{thimm} Theoretical predictions indicate that these
finite-size effects may  affect the screening of an artificial
impurity embedded  in an isolated  ring (i.e., not connected to
any external leads).\cite{affleck01}   It has indeed been proposed
that persistent currents in such a ring containing a quantum dot
may offer a way to probe directly this large Kondo length scale.
\cite{affleck01} ( See also Refs.
\onlinecite{hu,sorensen,eckle,kang,ferrari,aligia}.) When the
Kondo screening cloud becomes of order of the ring size or larger,
Ref. \onlinecite{affleck01} found a crossover in the
flux-dependence of the persistent current, from a (large
amplitude) saw tooth shape to a (small amplitude) sinusoidal
shape. Unfortunately, persistent current experiments are generally
delicate and very sensitive to disorder.

However, one may wonder what would be the signature of the
finite-size extension of the Kondo cloud when the ABI ring is
connected to several reservoirs. A priori, the physics is
different because the artificial impurity is now coupled to a
continuum and one may expect these finite-size effects to vanish.
In fact, this is not always the case, particularly when the
quantum dot is coupled to a finite-size mesoscopic wire, which is
itself weakly coupled to a reservoir.\cite{simon02,balseiro} It
has been shown that when the wire level spacing $\Delta_{r}$ is
larger than the reference Kondo temperature $T_K^0$ of a quantum
dot directly coupled to infinite leads, finite-size effects do
occur. The main consequence is a strong renormalization of the
genuine Kondo temperature of the artificial impurity, $T_K$, and a
strong dependence of $T_K$ on the local density of states seen by
the artificial impurity.\cite{simon02} In this paper we  therefore
analyze the transport properties of an ABI weakly coupled to
reservoirs and containing an in-line quantum dot tuned to be in
the Kondo regime. In particular we show that the Kondo temperature
strongly depends on the AB phase, leading to some direct
consequences on the finite temperature conductance through the
ABI. (For a discussion of the fluctuations of the Kondo
temperature as function of the magnetic flux in a chaotic ABI, see
Ref. \onlinecite{lewenkopf}.)

The plan of the paper is the following: In section \ref{sec:fse},
we describe our model Hamiltonian for the closed ABI and analyze
how finite-size effects affect the Kondo temperature. We
especially show that the Kondo temperature acquires a strong
dependence on the AB phase in a one-dimensional description of the
ABI. In section \ref{sec:conductance}, we present our results for
the conductance through the closed ABI using the slave-boson
mean-field theory approximation scheme.  In Sec. \ref{sec:openabi}
we analyze the Kondo temperature and the conductance  of an open
interferometer, by modifying our model to include numerous
additional leads. Finally, section \ref{sec:conclusion} contains a
discussion of our results and our conclusions. An appendix 
details the calculation of the conductance through the ABI
containing an interacting quantum dot.

\section{Energy scales in an AB interferometer}\label{sec:fse}

\subsection{Model Hamiltonian}

\begin{figure*}
\epsfig{figure=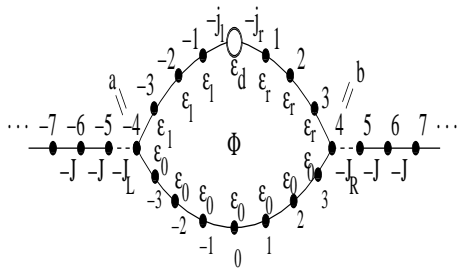,height=3.5cm,width=6cm} \caption{Our
model for the ABI with $n_l=n_r=4, n_0=2l_0+1=7$. The extremities
of the interferometer are denoted by $a$ and $b$. Here $a$ and $b$
correspond respectively to points $-4$ and $+4$.}
\label{Fig:interfero}
\end{figure*}

In order to analyze the role played by finite-size effects in an
ABI, we consider the simple one-dimensional model depicted in Fig.
\ref{Fig:interfero}. The tight-binding Hamiltonian describing the
ABI shown in that figure reads
\begin{equation}
\label{ham_abi} H=H_L+H_{I}+H_D+H_{L,I}+H_{I,D},
\end{equation}
where the subscripts $L$, $I$, and $D$ stand for the leads, the
interferometer, and the dot, respectively.
\begin{widetext}
\noindent In Eq. (\ref{ham_abi}),
\begin{eqnarray}
H_L&=&-J\left[
\sum_{i=-\infty}^{-n_l-2}+\sum_{i=n_r+1}^{\infty}\right]
\sum_{\sigma}(c^\dagger_{i\sigma}c^{\nil}_{i+1\sigma}+h.c.),\label{HLEAD}
\end{eqnarray}
is the lead Hamiltonian, and
\begin{eqnarray}
H_{I}&=&-J \sum_{i=-n_l}^{-2}
\sum_{\sigma}(c^\dagger_{i\sigma}c^{\nil}_{i+1\sigma}+h.c.)
+\epsilon_l
\sum_{i=-n_l}^{-1}\sum_{\sigma}n^{\nil}_{i\sigma}\nn\\
&&-J \sum_{i=1}^{n_r-1}
\sum_{\sigma}(c^\dagger_{i\sigma}c^{\nil}_{i+1\sigma}+h.c.)
+\epsilon_r \sum_{i=1}^{n_r}\sum_{\sigma}n^{\nil}_{i\sigma} -J
\sum_{i=-l_0-1}^{l_0}
\sum_{\sigma}(a^\dagger_{i\sigma}a^{\nil}_{i+1\sigma}+h.c.)
+\epsilon_0 \sum_{i=-l_0}^{l_0}\sum_{\sigma}a^\dagger_{i\sigma}
a^{\nil}_{i\sigma} , \label{HABI}
\end{eqnarray}
describes the ring, with
$n^{\nil}_{i\sigma}=c^{\dagger}_{i\sigma}c^{\nil}_{i\sigma}$. The
upper left and right branches of the ABI contain $n_l$ and $n_r$
sites, respectively, and  the lower arm contains $n_0=2l_0+1$
sites. In Eqs. (\ref{HLEAD}) and (\ref{HABI}),
$c^\dagger_{i\sigma}$ creates an electron with spin
$\sigma=\uparrow,\downarrow$ on site $i$ either in the upper
branch of the ABI (for $i\in[-n_l,n_r]$) or in the leads ($i<-n_l$
and $i>n_r$), whereas $a_{i\sigma}^\dag$ with $i\in[-l_0,l_0]$
creates an electron in the lower part of the ABI. (We use a
different labelling in order to distinguish electrons on the upper
branch from those on the lower branch.) We assign different sites
energies, $\eps_l$, $\eps_r$, and $\eps_0$ on these three
respective branches. We also identify $a_{-l_0-1}\equiv
c_{-n_l}\equiv c_a$ and $a_{l_0+1}\equiv c_{n_r}\equiv c_b$ (all
with spin $\sigma$) for the sites connecting the ring to the
leads. The Hamiltonian of the quantum dot is
\begin{eqnarray}
H_D&=&\epsilon_d\sum_{\sigma}n_{d\sigma}
+Un_{d\uparrow}n_{d\downarrow}.\label{HAMDOT}
\end{eqnarray}
In most of our calculations we assume that the Coulomb energy $U$
is larger than all other energies in the problem, and thus take
the limit $U \rightarrow \infty$. Also,
\begin{eqnarray}
&&H_{L,I}=-\sum_{\sigma}(J_{L}
c^\dagger_{-n_l-1\sigma}c^{\nil}_{-n_l\sigma}
+J_{R} c^\dagger_{n_r\sigma}c^{\nil}_{n_r+1\sigma}+h.c.),\nonumber \\
&&H_{I,D}=-e^{-i\al/2}\sum_{\sigma}(j_l
c^\dagger_{-1\sigma}c^{\nil}_{d\sigma}+j_r c^\dagger_{d\sigma}
c^{\nil}_{1\sigma})+h.c., \label{HLRING}
\end{eqnarray}
describe the coupling of the ring to the leads and to the dot. The
interferometer depicted in Fig. \ref{Fig:interfero} is threaded by
a magnetic field. This gives rise to the  phase factors appearing
in Eqs. (\ref{HLRING}), with $\alpha=2\pi \Phi/\Phi_0$, where
$\Phi$ is the total magnetic flux penetrating the ABI ring and
$\Phi_{0}$ is the flux quantum. Using gauge invariance, we have
distributed this phase on the tunneling amplitudes between the dot
and the ring.
 The total number of sites on the
interferometer, $L=n_{l}+n_{r}+n_{0}$, determines the size of the
ring.

In the following we will be interested in the regime where the
quantum dot describes the behavior of an artificial magnetic
impurity in the Kondo regime, i.e., when $n_d \equiv
n_{d\uparrow}+n_{d\downarrow} \sim 1$. The important energy scale
of the problem is then the Kondo temperature $T_K$. The next
section is devoted to an analysis of this energy scale.

\end{widetext}

\subsection{ The Kondo temperature}\label{sec:tk}

When the ring is disconnected from the reservoirs, it has been
shown in Ref. \onlinecite{affleck01} that the ring circumference
$L$ introduces a cutoff energy $\Delta_r=\hbar v_F/L$ (where
$v_{F}$ denotes the Fermi velocity of the conduction electrons),
which replaces the temperature $T$ in the renormalization of the
Kondo coupling when $T<\Delta_r$, hence preventing a perfect
screening of the impurity. The situation becomes more delicate
when finite-size effects (brought about by the finite length of
the ring) combine with the coupling to the continuum (described by
the leads). In Ref. \onlinecite{simon02}, a setup in which a
quantum dot is connected to finite-size wires which are in turn
weakly coupled to reservoirs has been analyzed. Then, the dot
non-interacting self-energy, and in particular the local density
of states (LDOS) on the dot,  become structured in energy due to
the finite-size effects. As a result, the Kondo temperature
depends on the fine structure of the LDOS.

In the present configuration, the finite-size ring is coupled to
the continuum, and in addition is threaded by a magnetic flux.
Indeed, we find below that the Kondo temperature varies strongly
(over several decades) with the magnetic flux when the ring energy
spacing $\Delta_r$ is larger than the bare Kondo temperature
$T_K^0$, or equivalently, when the ring length
 $L=\hbar v_F/\Delta_r$ is smaller
 than the Kondo length scale,
$\xi_K^0\equiv\hbar v_F/T_K^0$.  In fact, such a result can be
expected, at least qualitatively, as  can be understood by
considering special cases: In the limit where the ABI becomes
disconnected from the leads, we should recover the isolated ring
case studied in Ref. \onlinecite{affleck01}, where finite-size
effects were shown to appear when $T_K^0\ll \Delta_r$. Another
particularly interesting case is a symmetric (under parity) ABI
with only  a single site, $0$,  on the lower branch, and $n_l=n_r$
(in the notations used above). In this specific case, it is
convenient to perform a folding even/odd transformation by
defining $c^{e/o}_i=(c_i\pm c_{-i})/\sqrt{2}$, $\forall i>0$. In
the new basis, the quantum dot is coupled to two finite-size
wires,  of respective lengths $n_l+1$ and $n_l$, which are coupled
to a continuum via their other extremity. The AB phase enters only
through the tunnel amplitudes between the even/odd finite-size
wires and the dot: The corresponding matrix elements are
$j_l\sqrt{2}\cos(\al/2)$ for the even part and $i
j_l\sqrt{2}\sin(\al/2)$ for the odd part.\cite{note1}
 Such a situation was already studied
in Ref. \onlinecite{simon02},
 where it was shown that finite-size effects only
occurs for $n_l\lesssim \xi_K^0$.

On the other hand,  when the ring size is very large compared
to the Kondo length scale, finite-size effects are washed out.
 For
example, in the second particular case discussed in the previous
paragraph, it is easy to see that in the large-size ABI the Kondo
temperature is essentially flux-independent. \cite{note1}
On a more general footing, one may give the  following heuristic
argument: The
Kondo screening cloud is localized in the upper branch of the ABI
around the quantum dot. The AB phase can be assigned, by a gauge
transformation, to the lower branch alone. Then, electrons
participating in the dynamical screening of the impurity will not
``see'' the AB phase. In such a situation we expect the associated
Kondo temperature to be the bare Kondo temperature $T_K^0$.

From the aforementioned general arguments, one may therefore
expect that the Kondo temperature in an ABI will be affected both
by the finite size of the ABI branches  and by the magnetic flux.
The Kondo temperature is the energy scale separating the high
temperature perturbative regime, where the impurity is weakly
screened, from the low temperature strong-coupling regime, where
it forms a singlet with the conduction electrons. Since it is a
crossover scale,  there are many ways to define it, all capturing
the correct order of magnitude. In this paper, we mainly use
two
definitions, one resulting from the slave-boson mean-field theory  (see
Sec. \ref{sec:sbmft}), which expresses the Kondo temperature in
terms of the parameters of the non-interacting system (when there
are no interactions on the dot) and another one resulting from the
renormalization group approach.

The non-interacting  Green function on the dot can be written as
\beq
G_{dd}^0=[\omega-\epsilon_d-\Sigma_{dd}^0(\omega)]^{-1},\label{G0}
\eeq where
\begin{eqnarray}
\Sigma_{dd}^0(\omega)=\delta\epsilon(\omega)-i\Delta(\omega)\label{SE0}
\end{eqnarray}
is the dot non-interacting self-energy, and $\epsilon_{d}$ is the
single-particle on-site energy on the dot [see Eq.
(\ref{HAMDOT})]. For large $U$, the two parts of the
non-interacting self-energy determine {\it a priori} the Kondo
temperature. When the self-energy is a smooth function around the
Fermi energy, $E_{F}$, one obtains the  standard analytical
expression for the Kondo temperature,
\begin{eqnarray}\label{tkapprox}
\label{deftk} T_K\approx D_0\exp \left(\pi\frac{\epsilon_d
-\delta\epsilon(E_F)}{2\Delta(E_F)}\right),
\end{eqnarray}
where $D_0$ denotes the half-bandwidth on the leads. On the other
hand, when the self-energy  varies abruptly around the Fermi
energy (which is typically the case for the ABI), one needs to
solve numerically the self-consistent slave boson mean-field
integral equations in order to find that temperature. Denoting the
Kondo temperature of this situation by $T_K^{\rm SBMFT}$, one has
\begin{eqnarray}
T_K^{\rm SBMFT}\approx b_0^2\Delta(E_F),\label{tksbmft}
\end{eqnarray}
where $b_0$ is the slave boson parameter
(see Sec. \ref{sec:sbmft}).

Since the self-energy given in Eq. (\ref{SE0}) pertains to a
non-interacting system, it can be calculated in a straightforward
way. In the case of the Hamiltonian Eq. (\ref{ham_abi}) (see also
Fig. \ref{Fig:interfero}), it can be expressed as
\begin{eqnarray}
\label{sigmadd0} \Sigma_{dd}^0(\omega)=j_l^2 g_{-1-1}(\omega )
+j_r^2 g_{11}(\omega )+2j_lj_r\cos\alpha g_{-11}(\omega )
,\nonumber\\
\label{expSE}
\end{eqnarray}
where $g_{ij}(\omega )$ are the Green functions of the system
without the quantum dot (i.e., for $j_l=j_r=0$),  and therefore
refer to a non-interacting system.  Note that these Green
functions {\em do not} depend on the AB flux, so that the entire
flux dependence of the non-interacting self-energy comes from the
interference term in Eq. (\ref{expSE}).  This dependence makes
both $T_{K}$ and $T_{K}^{SBMFT}$ flux-dependent as well.

To ensure that this dependence is not a result of our definition
of the Kondo temperature, we have also computed the Kondo
temperature as defined by the renormalization group (RG)
technique, $T_{K}^{RG}$,\cite{hewson,simon02} \beq \label{tkrg}
\left[\int\limits_{-D_0}^{-T_K^{RG}}+\int\limits_{T_K^{RG}}^{D_0}\right]
d\omega \frac{J_{K}\rho(\omega)}{2|\omega|}=1, \eeq where $J_{K}$
is the Kondo coupling and $\rho(\omega)$ is the non-interacting
local density of states (LDOS) on the dot, given in our model by
$\Delta(\omega)/(j_l^2+j_{r}^{2})$. In the $U\to\infty$ limit of
the Anderson
 description of the quantum dot,
$J_{K}=2(j_l^2+j_r^2)/|\epsilon_d|$. It is worth noting that when
a quantum dot (in the Kondo regime) is connected to two large
reservoirs, $\rho(\omega)\sim\rho(E_F)$, and therefore
$T_K^{RG}=D_0\exp[-1/(J_{K}\rho(E_F)]=D_0\exp(\pi
\eps_d/2\Delta(E_F))$, which agrees with Eq. (\ref{deftk})
($\delta\epsilon(E_F)\sim 0$ in this case).

We exemplify the AB phase dependence of the Kondo temperature
below, utilizing two different parameterizations of the ABI.

\subsubsection{The Kondo temperature of a dot embedded in
a mesoscopic ABI ring}

As explained above, the quantity which determines the Kondo
temperature in the large-$U$ limit is the non-interacting
self-energy of the dot. Its calculation requires the Green
functions of the system without the dot. (Our Green functions are
the retarded ones, unless specified explicitly otherwise.)  One
may present those as a matrix of dimension $(L\times L)$ where
$L=n_{l}+n_{r}+n_{0}$ is the total number of sites on the
interferometer, with the dot excluded,
\begin{eqnarray}
\hat g=[\om\hat I-\hat H-\hat\Sigma]^{-1}. \label{ghat}
\end{eqnarray}
Here $\hat\Sigma$ is the self-energy due to the two (or more)
semi-infinite leads. In our model, $\hat\Sigma$ has just two
non-zero matrix elements, $\hat\Sigma_{aa}$ and $\hat\Sigma_{bb}$,
see Fig. \ref{Fig:interfero}. Once the matrix of Eq. (\ref{ghat})
is inverted, the result is used in Eq. (\ref{expSE}) to yield the
dot self-energy. Generally, this scheme is numerically quite time
consuming since for each frequency $\omega$ one needs to invert an
$L\times L$ matrix. To overcome this difficulty we utilize below
an approximate solution for a general ring, which exemplifies the
mesoscopic finite-size effects.  We then present an analytic
derivation for a very small model system where the ring contains a
single site besides the dot. Even this small ring, when weakly
coupled to the leads, already captures the strong dependence of
the Kondo temperature on the AB flux. However, such a small model
is insufficient to describe properly mesoscopic finite-size
effects.

\begin{figure}[t]
\psfig{figure=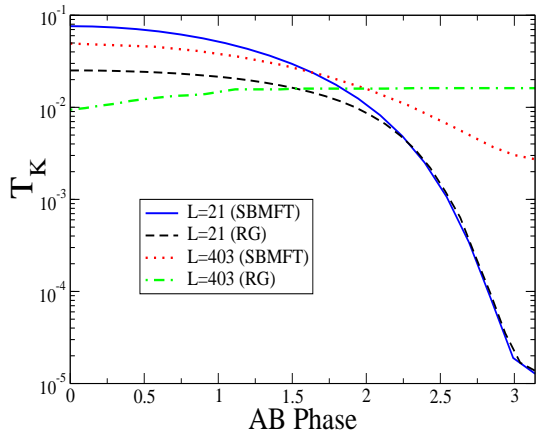,height=8.0cm,width=7.cm,angle=-90}
\caption{Kondo temperatures  as  function of the AB phase $\alpha
=2\pi \Phi/\Phi_{0}\in[0,\pi]$ calculated using either the SBMFT
definition,  Eq. (\ref{tksbmft}),  or the RG definition,  Eq.
(\ref{tkrg}), for two different ABI sizes: $L=21$ (the full  and
long dashed lines) and $L=403$ (the dotted  and dashed-dotted
lines). The following parameters are used: $j_l=j_r=0.35$,
$\eps_d=-0.7$ (giving $T_K^0\sim 0.011$ or $\xi_K^0\sim 180$)
 and $J_L=J_R=0.6$,
$\eps_l=\epsilon_{r}=\epsilon_{0}=0$.
 }
 \label{Fig:tk1}
\end{figure}

In the limit where the ABI is weakly-coupled to the leads (i.e.,
when $J_L^2,J_R^2\ll J^{2}$)  and for $\eps_0=\eps_l=\eps_r$, one
can obtain  a rather good approximation for the non-interacting
Green functions  by approximating the local density of states in
the ring by a Lorentzian. This approximation has been successfully
used previously in Ref. \onlinecite{simon02}.  In this
approximation,  the Green function $g_{ mn}$, where $m$ and $n$
are two sites on the ring,  is given by
\begin{eqnarray}
g_{mn}\approx\frac{2}{L+1}\sum\limits_{j=1}^L \frac{\sin(q_{rj}
m)\sin(q_{rj} n) } {\omega-\veps_j-\delta\eps_j+i
\Delta_j},\label{gappr}
\end{eqnarray}
where the variables $q$ and $\omega$ are related to one another by
$\omega=-2J\cos(q)$ (here and below we measure lengths in units of
the lattice constant). In Eq. (\ref{gappr}),
\begin{eqnarray}
\veps_j&=&-2J\cos(q_{rj})+\eps_r,
\end{eqnarray}
with
\begin{eqnarray}
q_{rj}&=&\frac{\pi j}{L+1}~~;~~j=1,\cdots,L .
\end{eqnarray}
For the case of two leads only, $\delta\varepsilon_{j}$ and
$\Delta_{j}$ are given by
\begin{eqnarray}
\delta\eps_j&=&-\frac{2}{L+1}\left(\frac{J_L^2}{J}
\sin^2(q_{rj} n_l)\sin(q)\right.
\nn\\&+&\left.
\frac{J_R^2}{J}\sin^2(q_{rj} n_r)\sin(q)\right),\nonumber\\
\Delta_j&=&\frac{2}{L+1}
\left(\frac{J_L^2}{J}\sin^2(q_{rj} n_l)\cos(q)\nn\right.\\
&+& \left. \frac{J_R^2}{J}\sin^2(q_{rj} n_r)\cos(q)\right).
\label{deltaapprox}
\end{eqnarray}

We have used the approximate expression (\ref{gappr}) for the
non-interacting Green functions, in conjunction with Eq.
(\ref{tksbmft}), to compute numerically the dependence of the
Kondo temperature, $T_K^{SBMFT}$, on the AB phase. Using
$n_l=n_{r}=5$, and $n_0=11$, which yield for the total size of the
ring $L=n_l+n_r+n_0=21$, one finds that the Kondo temperature
varies over at least three decades (the continuous line in Fig.
\ref{Fig:tk1}).  (The energy units are set by  the tight-binding
amplitude on the leads, $J=1$.)  This huge variation is not an
artifact of our definition of the Kondo temperature (resulting
from the slave-boson technique). Using the renormalization-group
definition, Eq. (\ref{tkrg}), we obtain, for the same parameters,
the long-dashed line in Fig. \ref{Fig:tk1}. Clearly, both
definitions give almost the same huge variation (differing around
$\alpha=0$ by a factor which is less than 3).

On the other hand, at larger ring sizes, the variations of the
Kondo temperature with the flux are strongly suppressed, as is
seen in Fig. \ref{Fig:tk1}.  We note that the RG definition
predicts  a Kondo temperature which is almost  phase-independent
(it varies by just a factor 2), whereas the SBMFT definition
predicts a variation by one decade. The reason for this difference
stems from the fact that the integral appearing in the RG
definition averages over the energy variations of the density of
states, making $T_K^{RG}$ less sensitive as compared to
$T_K^{SBMFT}$.\cite{note2} Nevertheless, the variations of $T_K$
of the $L=403$ ring are overall $\sim 3$ orders of magnitude
smaller than those of the $L=21$ one, proving that finite-size
effects are prominent for $L\lesssim \xi_K^0$. This strong
modulation of the Kondo temperature with the magnetic flux for
small ABI's will clearly affect the thermodynamic and transport
properties of a mesoscopic ring.

\subsubsection{The Kondo temperature of a smaller configuration}
\label{smallabi}

When the ring is very small, such that it contains a single site
in addition to the  dot, it is possible to find a simple
analytical expression for the non-interacting self-energy. Such a
system has been studied in  Ref. \onlinecite{ora04}. In our
parametrization, it  corresponds to a quantum dot directly
connected to the leads, and interfering with a lower path
characterized by a single energy level $\eps_0$ connected again
directly to the leads with hopping amplitudes $i_{\ell}$ and $i_r$
for the left and right leads, respectively. This model system
still demonstrates the remarkable variation of the Kondo temperature
with the AB phase.

In this model system the non-interacting dot self-energy is
\begin{eqnarray}\label{sigma0}
\Sigma^{0}_{dd}(\omega )=g_{L} (\omega
)(j_{\ell}^{2}+j_{r}^{2}+g^{0}_{L} (\omega )g_{00}(\omega
)|Y|^{2}).
\end{eqnarray}
Here, $g_{L} $ reflects the effect of the one-dimensional leads,
\begin{eqnarray}\label{gl}
g_{L} (\omega
)&=&\frac{2}{N}\sum_{k}\frac{\sin^{2}k}{\omega-\epsilon_{k}}
=-\frac{e^{iq}}{J},
\end{eqnarray}
$g_{00}$ is the Green function of the
lower arm of the interferometer, when disconnected from the dot,
\begin{eqnarray}\label{g00}
g_{00}(\omega )&=&\frac{1}{\omega -\epsilon_{0}-g_{L} (\omega
)(i_{\ell}^{2}+i_{r}^{2})},
\end{eqnarray}
 and
\begin{eqnarray}\label{Y2}
|Y|^{2}=j_{\ell}^{2}i_{\ell}^{2}+
j_{r}^{2}i_{r}^{2}+2j_{\ell}j_{r}i_{\ell}i_{r}\cos\alpha ,
\end{eqnarray}
containing the dependence of the self-energy on the AB flux.

\begin{figure}[t]
\psfig{figure=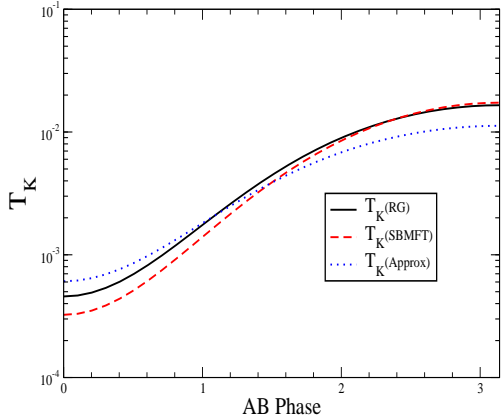,height=7.50cm,width=7.cm,angle=-90}
\caption{The Kondo temperature as  function of the AB phase,
$\alpha =2\pi\Phi/\Phi_{0}\in[0,\pi]$, for the small configuration
described in Sec. \ref{smallabi}. The parameters used are :
$T_{B}=0.5$, $\epsilon_{d}=-0.7$,
$\epsilon_l=\epsilon_r=\epsilon_{0}=-1$ and $j_l=j_r=0.35$.}
\label{Fig:tk2}
\end{figure}

To use these results in the expression for the Kondo temperature,
Eq. (\ref{deftk}), we take the Fermi energy to be in the middle of
the energy bands of the leads, i.e., $\omega =0$ and $q=\pi /2$.
Then
\begin{eqnarray}
g_{00}(E_{F})&=&\frac{1}{-\epsilon_{0}+
i\frac{i_{\ell}^{2}+i_{r}^{2}}{J}} .
\end{eqnarray}
For a symmetric configuration, $i_{\ell}=i_{r}$ and
$j_{\ell}=j_{r}$, the self-energy at zero frequency can be written
in terms of the transmission
($T_{B}=4(i_{\ell}^{2}/J)^{2}/[\epsilon_{0}^{2}+4(i_{\ell}^{2}/J)^{2}]$)
and reflection ($R_{B}=1-T_{B}$) of the lower arm of the ring,
yielding
\begin{eqnarray}
\Delta (E_{F})
&=&2\frac{j_{\ell}^{2}}{J}(1-T_{B}\cos^{2}\frac{\alpha}{2}),\nonumber\\
\delta\epsilon (E_{F})
 &=&{\rm sgn}(\epsilon_{0})2
\frac{j_{\ell}^{2}}{J}\sqrt{R_{B}T_{B}}\cos^{2}\frac{\alpha}{2}.
\end{eqnarray}
The dependence of the Kondo temperature on the AB phase,
calculated for this model using either the RG definition, Eq.
(\ref{tkrg}), or the SBMFT definition, Eq. (\ref{tksbmft}), or the
approximate definition, Eq. (\ref{tkapprox}), is depicted in Fig.
\ref{Fig:tk2}. It is seen that even for this minimal geometry, the
Kondo temperature varies by a factor $\sim 50$. By taking smaller
values of the dot tunneling amplitudes, this factor can be even
further  enhanced by several orders of magnitude.

\section{The conductance through the ABI}\label{sec:conductance}

\subsection{The slave-boson mean-field theory}\label{sec:sbmft}

In order to study the finite temperature transport through the
ABI, we need a reliable approximation scheme that will describe
well the electronic correlations on the dot localized level on one
hand, and will be sufficiently easy to handle on the other. Since
the conductance through the ABI can be expressed in terms of the
dot Green function,  an appropriate approximation scheme should be
applied to this entity. We choose to employ the slave-boson
mean-field theory (SBMFT). This technique is known to capture
qualitatively the low-temperature properties of the Anderson
model, and in particular it  provides a correct estimate of the
Kondo temperature,  as discussed above.

For the application of the slave-boson technique  to the ABI, it
is useful to re-write the model Hamiltonian Eq. (\ref{ham_abi}) in
a way that singles out the parts which contain the dot operators.
This is accomplished by writing
\begin{eqnarray}
H=H_{D}+H_{tun}+H_{net},\label{HSB}
\end{eqnarray}
where $H_{D}$ is the dot Hamiltonian,  given in Eq.
(\ref{HAMDOT}), $H_{tun}$ describes the coupling between the dot
and the ABI and is given by the second equation of (\ref{HLRING}),
and $H_{net}$ contains all other parts of the Hamiltonian (which
are non-interacting).

The slave-boson technique is usually applied \cite{hewson} in the
case where the Hubbard $U$ is the largest energy of the problem.
In practice, we set $U=\infty$. In that limit, the creation
operator on the quantum dot can be written as
$c^{\dagger}_{d\sigma}=f^\dag_\s b$, where the fermionic operator
$f^{\dagger}_{\sigma}$ creates a singly occupied state on the dot,
while the bosonic operator $b$ creates an empty state (i.e., a
two-hole state) there. Since $U=\infty$, the dot can be at most
singly occupied, and therefore one must impose the constraint
\begin{eqnarray}
b^\dag b+\sum\limits_\s f^\dag_\s f_\s=1. \label{constraint}
\end{eqnarray}

 In the mean-field treatment of the slave-boson technique,
the boson operator $b$ is replaced by a c-number, $b_{0}$, and the
constraint (\ref{constraint}) is implemented by introducing the
Lagrange multiplier $\lambda_{0}$. In this way, the Hamiltonian
becomes a non-interacting one, and is consequently easy to solve.
However, the parameters $b_{0}$ and $\lambda_{0}$ have to be
solved self-consistently.

Applying the SBMFT to our model Hamiltonian Eq. (\ref{HSB}), we
find that the part referring to the dot alone changes into
\begin{eqnarray}
H_{D}\to \epsilon_f \sum_{\sigma}f^\dag_{\sigma}
f^{\nil}_{\sigma}+\lam_0(b_0^2-1),\label{hammf1}
\end{eqnarray}
where we have defined
$\eps_f=\eps_d+\lam_0$, and $H_{tun}$ changes into
\begin{eqnarray}
H_{tun}=-e^{-i\alpha
/2}b_{0}\sum_{\sigma}(j_{\ell}c^{\dagger}_{-1\sigma}f^{\nil}_{\sigma}
+j_{r}f^{\dagger}_{\sigma}c^{\nil}_{1\sigma})+h.c.\ .
\label{hammf2}
\end{eqnarray}
The values of $\lam_0$ and $b_0$ are determined  by minimizing the
free energy of the system, defined by $F_{MF}=-{1\over \beta}\log
Z+\lam_0(b_0^2-1)$, where $Z$ is the partition function. The
mean-field free energy is conveniently expressed in the form
\begin{eqnarray}
F_{MF}=-{2\over \pi}\int\limits_{-D_0}^{D_0} d\om~ f(\om) \imag
[\ln G_f^R(\om)]+\lam_0(b_0^2-1) ,\label{FSB}
\end{eqnarray}
where $f(\om)=1/[1+\exp(\beta\om)]$  is the Fermi function, and
$G_f^R$ is the retarded Green function on the dot. Since the
mean-field Hamiltonian is non-interacting, this retarded Green
function is given by
\begin{eqnarray}
G_f^R(\om)=[\om+i\eta-\eps_f-\Sigma^R_{f}(\om)]^{-1},\label{GSB}
\end{eqnarray}
with $\Sigma_{f}^{R}(\om)=\delta\eps_{f}(\om)-i\Delta_{f}(\om)$
being the dot self-energy associated with the SBMFT, Eqs.
(\ref{hammf1}) and (\ref{hammf2}). This self-energy, in turn, is
given by our original non-interacting self-energy, [see Eq.
(\ref{G0})] $\Sigma_{f}^{R}(\om)=b_0^2 \Sigma^{0R}_{dd}(\om)
=b_0^2[\delta\eps(\om)-i\Delta(\om)]$, where
$\Sigma_{dd}^{0R}(\om)$ does not depend on $b_0$.

Inserting Eq. (\ref{GSB}) into Eq. (\ref{FSB}), the  mean-field
free energy becomes
\begin{eqnarray}
F_{MF}=&&{2\over \pi}\int\limits_{-D_0}^{D_0} d\om~ f(\om)
\arctan\left({b_0^2 \Delta(\om)\over
\om-\eps_f-b_0^2\delta\eps(\om)}\right)\nn\\
+&&(\eps_f-\eps_d)(b_0^2-1) .
\end{eqnarray}
Minimizing it with respect to the parameters  $b_0$ and $\eps_f$
(which replaces the parameter $\lambda_{0}$), leads to a set of
two coupled integral equations,
\begin{eqnarray}
&&{2\over \pi}\int\limits_{-D_0}^{D_0} d\om~ f(\om)
{\Delta_{f}(\om)\over
[\om-\eps_f-\delta\eps_{f}(\om)]^2+\Delta_f^2(\om) }\nn\\&&
  +b_0^2-1=0,\nonumber\\
&&{2\over \pi}\int\limits_{-D_0}^{D_0} d\om~ f(\om)
{\Delta(\om)(\om-\eps_f)\over
[\om-\eps_f-\delta\eps_{f}(\om)]^2+\Delta_{f}^2(\om) }\nn\\
&&+(\eps_f-\eps_d)=0 .\label{SBMFTEQ}
\end{eqnarray}
In the simplest case in which the self-energy is independent of
the frequency,  $\delta\epsilon_{f}(\om)\to b_0^2\delta\epsilon$
and $\Delta_{f}(\om)\to b_0^2\Delta$, these two equations can be
solved in a straightforward manner at zero
temperature.\cite{hewson} In the general case they must be solved
numerically by an iteration procedure, as function of the
temperature. In this way one obtains the dot Green function, which
is used below in the calculation of the conductance.

\subsection{SBMFT results for the conductance
of a weakly-coupled closed ABI}\label{sbmft:conductance}

\begin{figure}[h]
\psfig{figure=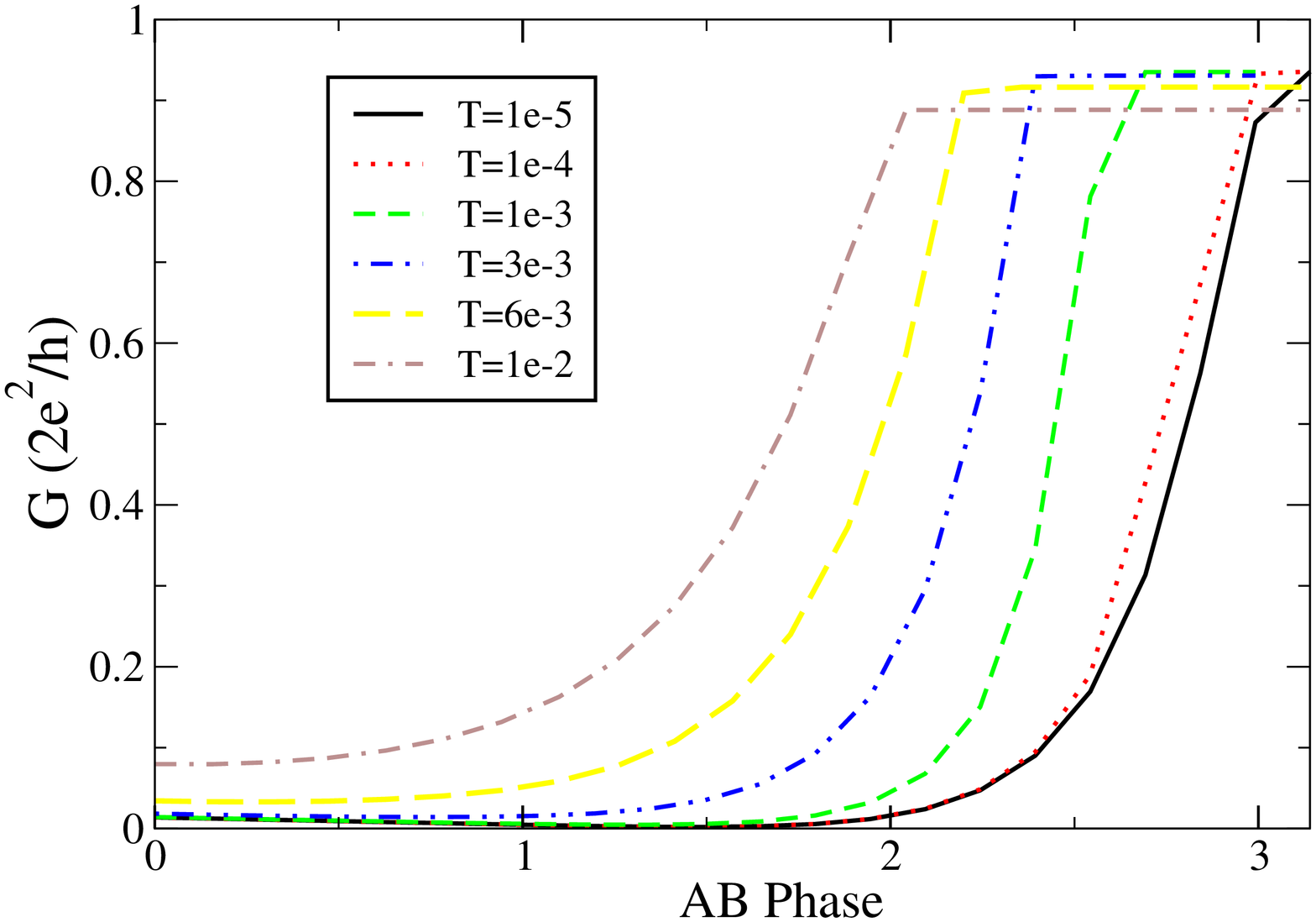,height=8.6cm,width=7.5cm,angle=-90}
\psfig{figure=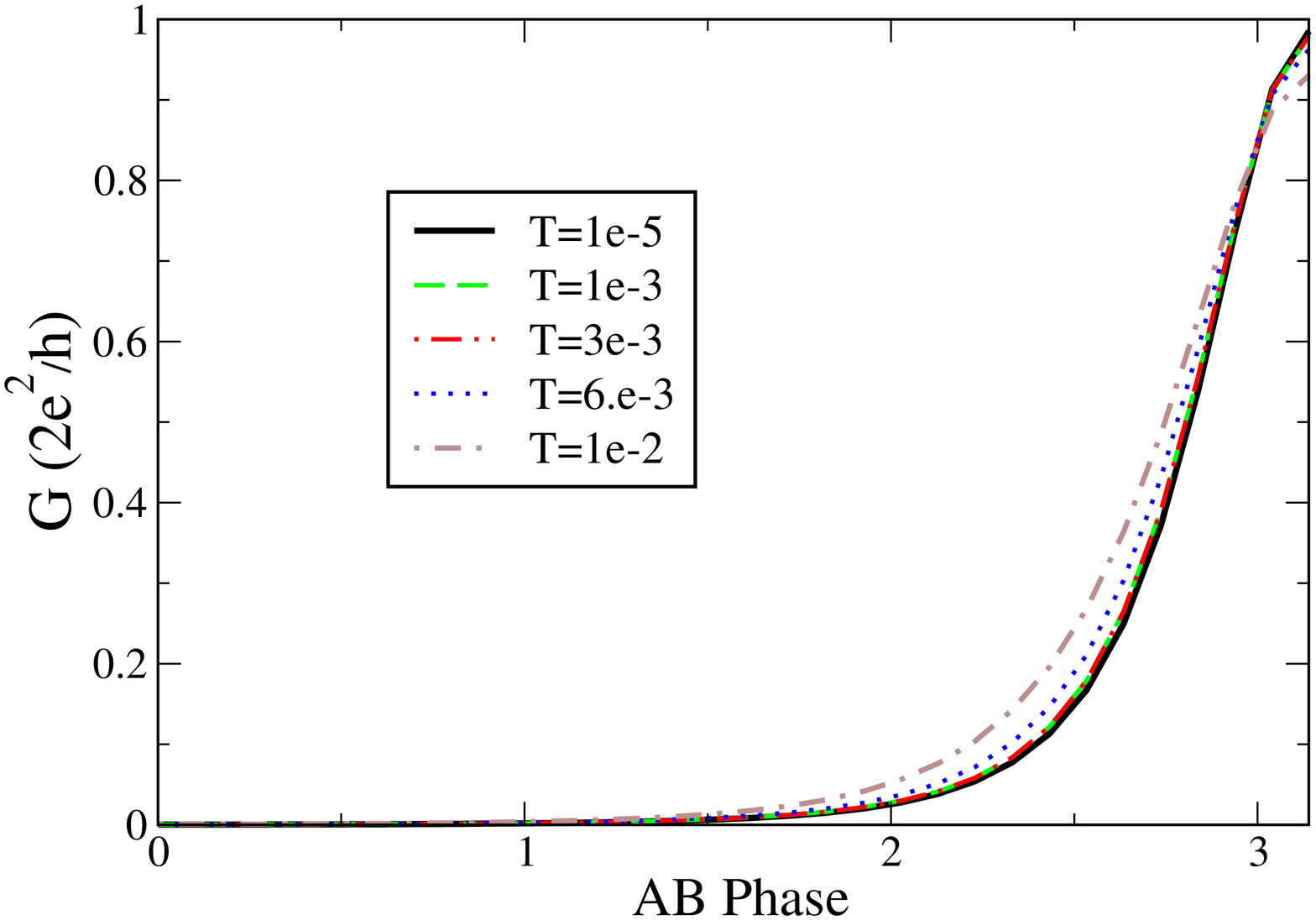,height=8.6cm,width=7.5cm,angle=-90}
\caption{The conductance as  function of the AB phase,
$\alpha=2\pi\Phi/\Phi_0$, for various values of the temperature.
The upper panel shows the conductance when the dot is in the Kondo
regime, and the lower panel shows it for a  non-interacting
quantum dot at resonance.} \label{Fig:conductance}
\end{figure}

A general expression for the finite temperature conductance
through the ABI can be derived as function of the exact retarded
Green function on the dot and the ABI parameters. This expression
is derived in the Appendix. However, it is worth noting that since
in the SBMFT the Green function on the dot is that of a
non-interacting system, the resulting expression is equivalent to
that derived from the Landauer-like formula,\cite{caroli} \beq
G=\frac{2e^2}{h} \int d\omega
 \frac{4 J_L^2J_R^2(1-\frac{\omega^2}{4J^2})}{J^2}
 \left(\frac{-\partial f}{\partial \omega} \right)
|G_{ab}^R|^2,\label{LANDAUER} \eeq where $G_{ab}^R$ is the exact
retarded Green function for the sites $a$ and $b$.

Using the expression (\ref{LANDAUER}), we have calculated the
conductance as function of $\alpha\in[0,\pi]$ for several
temperatures, using parameters as in section \ref{sec:tk},
 $j_l=j_r=0.35$, $\eps_d=-0.7$, $J_L=J_R=0.6$, $n_l=n_r=5$,
$n_0=11$, and $\epsilon_l=\epsilon_r=\eps_0=0$. These calculations
have been performed employing the approximation of weakly coupled
leads, as summarized by Eqs. (\ref{gappr})-(\ref{deltaapprox}).
The results are depicted in Fig. \ref{Fig:conductance}: The upper
panel pertains to the situation in which the quantum dot is in the
Kondo regime, and the lower panel is for a non-interacting quantum
dot at resonance.  Note that the conductance is an even function
of the AB phase $\alpha$ and has a $2\pi$ periodicity.

When the temperature is below the Kondo temperature $T_K(\alpha)$,
 for any value of  the AB phase $\alpha$, the spin of the
artificial impurity is almost fully screened.  This situation
corresponds to the unitary limit of the conductance. In this
temperature regime, the quantum dot is perfectly transmitting and
can be effectively replaced by a non-interacting quantum dot at
resonance. Indeed, the conductances displayed in the upper and
lower panels of Fig. \ref{Fig:conductance} are comparable for
$T=10^{-5}$. However, when the temperature becomes of order of the
Kondo temperature or larger, the transmission of the interacting
dot is strongly suppressed. Therefore, transport through the ABI
is essentially carried via the lower branch, and we thus expect
the conductance to be flux independent. Due to the huge variation
of the Kondo temperature with the AB phase $\alpha$, the
interesting situation occurs when $\min[T_K(\alpha)]\ll T\ll
\max[T_K(\alpha)]$. In this temperature regime, there are domains
of $\alpha$ in which the dot transmission is very small; these are
associated with plateaus of a large conductance through the ABI.
This is illustrated in the upper panel of Fig.
\ref{Fig:conductance}, where such plateaus appear around
$\alpha=\pi$ (for our choice of parameters, this regime
corresponds to the lowest Kondo temperature) and become larger as
we increase the temperature. The more we increase the temperature,
the more dramatic this effect is. Strictly speaking, when the
temperature is increased beyond the Kondo temperature, the SBMFT
approximation loses its validity. Nevertheless, we believe that
the main effect, i.e., the strong decrease of the dot
transmission, is qualitatively captured.  It seems plausible that
within a more numerically-accurate approach like the numerical
renormalization group (NRG), the conductance curves will be
smoothed, such that there will  be a weak phase dependence in the
plateau regime. Such a weak dependence cannot be captured by the
present method.

In order to highlight that these features are associated with
interaction effects, we also plot the conductance at  the same
temperatures for a non-interacting quantum dot (see the lower
panel of Fig. \ref{Fig:conductance}). The curves are very similar
to the ones in the upper panel of Fig. \ref{Fig:conductance} when
the temperature is much smaller than the (minimal) Kondo
temperature, but  differ drastically at some intermediate
temperature. The fact that the conductance does not completely
reach the unitary limit even for the non-interacting situation is
due to the approximations [Eqs. (\ref{gappr})-(\ref{deltaapprox})]
used in calculating the non-interacting Green functions. This
inability to reach the unitary limit is actually enhanced by the
SBMFT approximation for the interacting case (see the upper panel
of Fig. \ref{Fig:conductance}). Nevertheless,
 the main features which we describe
are qualitatively well reproduced.

\vspace{1cm}
\begin{figure}[h]
\psfig{figure=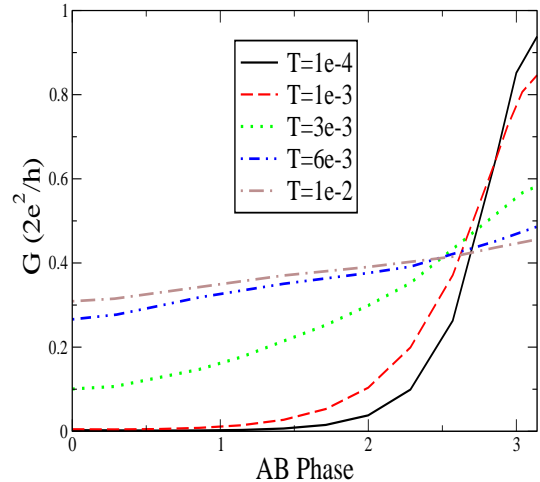,height=8.cm,width=8.cm,angle=-90}
\caption{The conductance as function of the AB phase
$\alpha=2\pi\Phi/\Phi_0$ for various values of the temperature,
for $L=403$ ($n_l=n_r=101$, $n_0=203$). The other parameters are
unchanged. The dot is  in the Kondo regime.
} \label{Fig:g_large}
\end{figure}

 The qualitative features of the results depicted in Fig.
 \ref{Fig:conductance}  do not depend on our particular choice of the
parameters. They reflect  the strong modulation of the Kondo
temperature with the flux, and are associated mainly with the
finite size of the ring. In order to further exemplify this point
we have calculated the variation of the conductance at different
temperatures for a large-size ABI, having $L=403>\xi_K^0$. The
results are depicted in Fig. \ref{Fig:g_large}.  At this large
size the slave-boson parameters, $b_0$ and $\eps_f$, are almost
independent of the AB flux, eliminating effectively the
flux-dependence of the interaction effects. Then, the flux
dependence of the ABI properties arise simply from the flux
dependence of the non-interacting self-energy. \cite{note2} As a
result, the conductance no longer displays the plateau-like
features that appeared for the small-size ABI. It is only at high
temperatures, $T\geq 6 \cdot 10^{-3}$, that the conductance
appears flatter almost over the entire flux range. The conductance
curves in Fig. \ref{Fig:g_large} can be well interpreted by a
Kondo temperature which is almost independent of the flux (as
corroborated by Fig. \ref{Fig:tk1}). At temperatures higher than
the Kondo one, the transmission through the dot becomes small,
independently of the AB flux, and the conductance is mainly
through the lower arm of the interferometer.

\subsection{SBMFT results for the conductance through the small ABI}

\begin{figure}[h]
\epsfig{figure=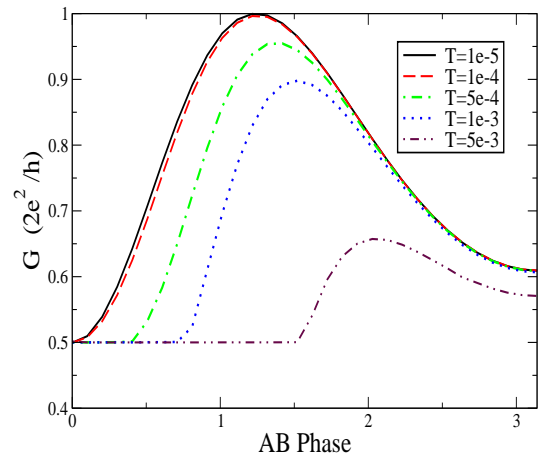,height=8.cm,width=7.5cm,angle=-90}
\caption{The conductance of the smallest ABI configuration, as
function of the AB phase $\alpha=2\pi\Phi/\Phi_0$ for various
values of the temperature. } \label{Fig:g_smallabi}
\end{figure}

 In is interesting to use the slave-boson technique to study
the small ABI configuration, introduced in Sec. \ref{smallabi}
above. Since within the SBMFT one effectively deals with a
non-interacting system, the conductance is given, via the Landauer
formula, by the transmission [with the various dot parameters
being computed self-consistently from the SBMFT Eqs.
(\ref{SBMFTEQ})]. In the case of the small configuration, it is
easy to obtain an explicit expression  for the energy-dependent
transmission coefficient, ${\cal T}(\omega )$,
\begin{eqnarray}
&&{\cal T}(\omega )=\frac{4(1-\frac{\omega^{2}}{4J^{2}})}{J^{2}}
\Big |g_{00}(\omega )i_{\ell}i_{r}\Bigl (1+\Sigma_{f}^{R}(\omega
)G^{R}_{f}(\omega )\Bigr )\nonumber\\ &+& G^{R}_{f}(\omega
)b_{0}^{2}j_{\ell}j_{r}e^{-i\alpha}\Bigl (1+g_{L}(\omega
)g_{00}(\omega )(i_{\ell}^{2}+i_{r}^{2})\Bigr )\Big
|^{2}.\label{trans}
\end{eqnarray}
Here, the self-energy $\Sigma_{f}^{R}$ is given by multiplying the
terms of Eq.  (\ref{sigma0}) by $b^{2}_{0}$. Note that ${\cal
T}(\omega )$ is even in $\alpha$, and hence satisfies the Onsager
relations. This can be seen by factoring out from both members in
$|\ldots |^{2}$ of Eq. (\ref{trans}) the product
$G^{R}_{f}g_{00}$. The remaining terms are real except for the
factor $e^{-i\alpha}$.

The  conductance through the small ABI  as function of the AB
phase $\alpha$, at various temperatures, is  depicted in Fig.
\ref{Fig:g_smallabi}, for the same parameters as used in
calculating the Kondo temperature (see  Fig. \ref{Fig:tk2}). At
low temperatures, $T\leq 10^{-4}$, the curve is smooth. It
corresponds to the low temperature regime $T\ll T_K(\alpha)$ in
which the spin impurity is well screened at all values of the flux
(the so-called unitary limit). At higher temperatures there appear
the plateau-like features around $\al=0$. These values of the flux
correspond to the lowest Kondo temperatures (see Fig.
\ref{Fig:tk2}), where the dot has a small transmission for $T\gg
T_K(\alpha)$. It is remarkable that even this small ABI
configuration already displays the main features associated with
the flux-dependent Kondo temperature.  Note that the plateaus
appear here at the smallest values of the conductance (as opposed
to the previous case, see Fig. \ref{Fig:conductance}). This
feature simply reflects a non-interacting interference effect that
depends on the specific choice of parameters which determine the
detailed dependence of the Kondo temperature on the AB flux.

\section{The open ABI}\label{sec:openabi}

 As is mentioned in the Introduction, there is much interest in
the complex transmission amplitude,
$t_{QD}=|t_{QD}|e^{i\varphi_{QD}}$, of a quantum dot, particularly
in the Kondo regime. While $|t_{QD}|$ can be inferred from the
conductance of the quantum dot, when coupled to two leads, this is
not the case with the phase shift $\varphi_{QD}$. It has been
conjectured that open ABI's, under certain conditions, will mimic
the two-slit limit. Has it been the case, the total transmission
amplitude of the interferometer threaded by an AB flux would have
read $t_{ABI}\propto t_{QD}e^{i\alpha}+t_{B}$, where $t_{B}$ is
the transmission amplitude of the other arm of the ring. Then, by
varying the quantum dot properties (for example, the gate voltage
on it, which in our model is $\epsilon_{d}$), one would have been
able to record the phase shift $\varphi_{QD}$ by monitoring the
variations of the conductance (which is determined by
$|t_{ABI}|^{2}$) with the AB flux. \cite{ji} Unfortunately, there
are several caveats in this attractive scenario. Firstly, the open
ABI has to obey rather stringent conditions in order to be in the
two-slit limit. \cite{schiller,abh} Secondly, when the quantum dot
is placed on the interferometer, its properties are in general
changed, because of its coupling to the mesoscopic ring, and due
to the presence of the AB flux. In the following, we do not
attempt to search the parameter space in order to find the regions
where the open ABI is in the two-slit limit. Rather, we first
study the Kondo temperature of an open ABI, and then examine its
conductance, as function of the flux, for several values of the
gate voltage.

In order to `open' the ABI in our tight-binding description (see
Fig. \ref{Fig:interfero}), we attach to each site inside the ring
(excluding the sites $a$ and $b$ which are already attached to
leads) a lead connected in turn to an electronic reservoir,
assigning a hopping matrix element $-J_X$ to its first bond. Such
parametrization has been previously used in Ref.
 \onlinecite{abh} to study the
conductance through a non-interacting open ABI. Technically, one
uses Eq. (\ref{gappr}) again, but now Eq. (\ref{deltaapprox}) is
replaced by a sum of contributions from all the connections to the
additional leads.

\begin{figure}[h]
\psfig{figure=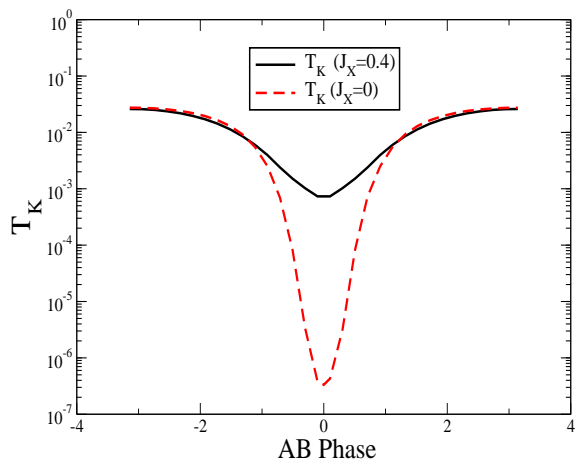,height=8.5cm,width=7.5cm,angle=-90}
\caption{The Kondo temperature of an open ABI as  function of the
AB phase for $J_X=0.4$ (full line) and $J_X=0$ (dashed line). The
other parameters used are: $j_l=j_r=0.35$, $\eps_d=-0.7$,
$n_l=n_r=3$, $n_0=5$, $J_L=J_R=0.6$. } \label{Fig:tk_open}
\end{figure}

\begin{figure}[h]
\psfig{figure=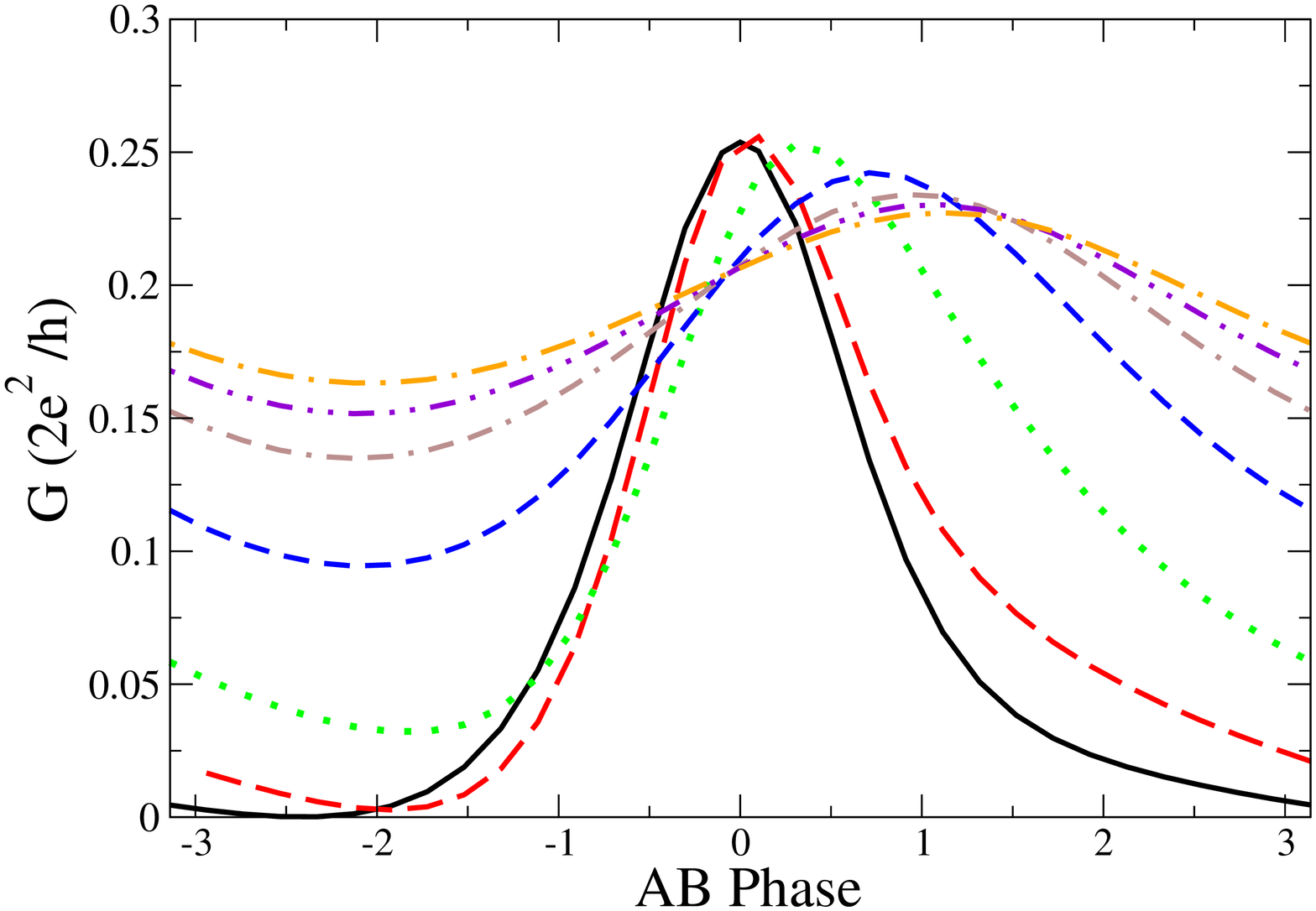,height=8.5cm,width=7.5cm,angle=-90}
\psfig{figure=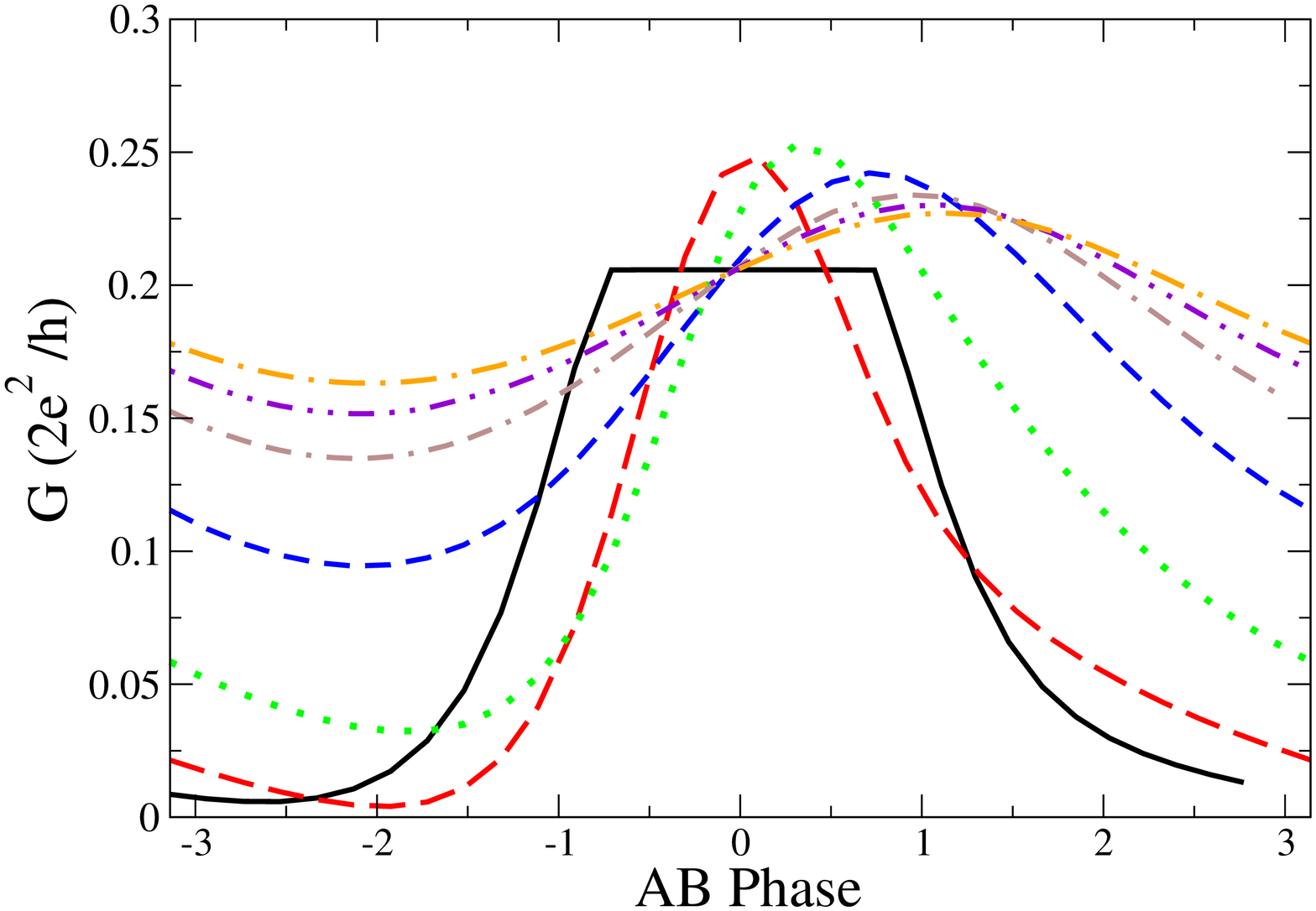,height=8.5cm,width=7.5cm,angle=-90}
\caption{The conductance of the open ABI as function of the AB
phase for $T=10^{-4}$ (upper panel) and $T=3 \cdot 10^{-3}$ (lower
panel) for different values of $\eps_d$: $\eps_d=-0.7$ (full
line), $\eps_d=-0.5$ (long-dashed line), $\eps_d=-0.3$ (dotted
line), $\eps_d=0$ (short-dashed line), $\eps_d=0.3$
(dashed-dashed-dotted line), $\eps_d=0.5$ (dotted-dotted-dashed
line), and $\eps_d=0.7$ (long dashed-dotted line). }
\label{Fig:g_open}
\end{figure}

Figure \ref{Fig:tk_open} depicts the dependence of the Kondo
temperature of an open ABI on the flux, when the length of the
ring is $L=11\ll \xi_K^0\sim 180$, taking $J_X=0.4$.  For
comparison, we have also plotted the Kondo temperature of the
closed ABI (for which $J_X=0$) for the same parameters. We have
used the RG definition, Eq. (\ref{tkrg}), to produce these curves
[we have verified that Eq. (\ref{tksbmft}) gives similar results].
It is interesting to note that the bare Kondo temperature
pertaining to these parameters is $T_K^0\sim 0.011$. Namely, the
Kondo temperature of a dot placed on an ABI is significantly
different from the bare one. As can be seen in Fig.
\ref{Fig:tk_open}, the dependence of the Kondo temperature on the
AB flux of the open ABI is considerably suppressed as compared to
that of the closed one (by more than 3 orders of magnitude!). This
implies that the open ABI is less sensitive to finite-size effects
than the closed ABI, as could be anticipated: `Opening' the
interferometer smoothes the fine structure of the non-interacting
self-energy, and consequently reduces the dependence of the Kondo
temperature on the AB phase. Nonetheless, finite-size effects in
the open ABI have not completely disappeared, and the Kondo
temperature still varies by a factor $\sim 50$.

 Next we study the conductance through the open ABI. For such
an interferometer, the conductance is no longer an even function
of the AB flux. As is discussed above, one is mainly interested in
the manner by which the deviation from an even function is
modified when the gate voltage, i.e., $\epsilon_{d}$, is varied.
Using the slave-boson technique, we have computed the conductance
for several values of this parameter. Figure \ref{Fig:g_open}
shows it at two temperatures, $T=10^{-4}\ll T_K$ (upper panel) and
$T=3 \cdot 10^{-3}$ (lower panel). At the lower temperature, the
conductance evolves smoothly, and it is possible to extract a
phase shift associated with the shift of the maximum of the
conductance upon varying $\eps_d$. For example, going from the
deep Kondo regime ($\eps_d=-0.7$, $n_d\sim 1$) to the empty regime
($n_d\sim 0$, $\eps_d=+0.7$), this phase shift corresponds here to
$\sim 0.35 \pi$. However, it strongly depends on the other
parameters of the ABI (as has been explicitly shown in Ref.
\onlinecite{aharony04}). At the higher temperature the shift
becomes more difficult to read since the maximum around $\al=0$
for $\eps_d=-0.7$ (deep Kondo regime) is now replaced by a
plateau.

\section{Conclusion}\label{sec:conclusion}

In this paper we have used the SBMFT to calculate the
flux-dependence of both the Kondo temperature of a quantum dot
embedded on one branch of an ABI and the conductance through the
ABI. One should be aware the SBMFT is not suitable to describe
properly the dot charge fluctuations. Therefore, it poorly
describes the mixed valence regime and loses its validity in the
empty-dot regime. Also, since the slave-boson technique assumes
the $U\to \infty $ limit, it cannot capture the resonance
transmission faithfully, and it may fail in  mimicking the details
of the experiments. Nonetheless, the above results, particularly
in the Kondo regime, demonstrate some qualitative features, which
we expect to also arise in more accurate calculations.

Our calculations yields the following conclusions:

\begin{itemize}
\item For both the closed and the open ABI, the Kondo temperature
of the dot depends on the magnetic flux penetrating the ABI ring.
This dependence is very strong (several orders of magnitude) for
small ABI's, and weaker for larger ABI's. Thus, the effects of
electron interactions strongly depend on the size of the ABI.

\item As a result of the above, the flux-dependence of the
conductance through the ABI also varies with the size. For small
sizes, the system moves from below to above the Kondo temperature
as function of the flux, yielding plateaus in the conductance.
These plateaus broaden at higher temperature. Unless the
temperature is much smaller than the minimal $T_K(\alpha)$, one
cannot expect to fit the experimental data to any universal
function $G(\alpha)$.

\item The above conclusions apply to both the closed and the open
ABI. In particular, they imply that it is very difficult to
construct an open ABI that will obey the two-slit formula. In most
cases, the shift of the maxima in $G(\alpha)$ with the gate
voltage (as represented by $\epsilon_d$) does {\it not} reflect
the corresponding shift in the transmission phase through the bare
dot. \cite{physica}

\end{itemize}

\acknowledgments

This project was carried out in a center of excellence supported
by the ISF under grant No. 1566/04. Work at Argonne is supported
by  the U.S. Department of Energy under contract W-31-109-Eng-38.
Part of the numerical calculations presented in this paper were performed
on the cluster MedEtPhy (CIMENT, Grenoble).
\vspace{1cm}

\appendix
\section{A general expression
for the current through an Aharonov-Bohm interferometer}

Here we use the Keldysh technique to derive a general expression
for the current through an ABI described by the Hamiltonian Eq.
(\ref{ham_abi}) (see also Fig. \ref{Fig:interfero}).

We begin by writing down the expressions for the currents
$I_{\ell}$ and $I_{r}$, entering the ABI from the left lead and
from the right lead, respectively,
\begin{eqnarray}
I_{\ell}&=&-\frac{e}{h}\int d\omega
[\Sigma_{L,\ell}(\omega )G_{n_{\ell}n_{\ell}}(\omega )
-G_{n_{\ell}n_{\ell}}(\omega )\Sigma_{L,\ell}(\omega )]^{<},\nonumber\\
I_{r}&=&-\frac{e}{h}\int d\omega [\Sigma_{L,r}(\omega
)G_{n_{r}n_{r}}(\omega )-G_{n_{r}n_{r}}(\omega
)\Sigma_{L,r}(\omega )]^{<}. \nonumber\\
\label{IlIr}
\end{eqnarray}
As in the main text, we denote by $G$ the Green functions of the
full ABI, and by $g$ the Green functions of the ABI when the dot
is disconnected from the ring. In Eq. (\ref{IlIr}), the
superscript $<$ denotes the lesser Keldysh Green function, with
the lesser Green function of a product being calculated according
to the rules derived in Ref. \onlinecite{langreth1}. $\Sigma_{L}$
is the self-energy due to the semi-infinite leads,
\begin{eqnarray}
[\Sigma_{L}]_{ij}(\omega )=\delta_{ij}(\delta_{i,-n_{\ell}}
\Sigma_{L,\ell}(\omega )+\delta_{i,n_{r}}\Sigma_{L,r}(\omega )).
\end{eqnarray}
Here $\Sigma_{L,\ell}$ and $\Sigma_{L,r}$ are the self-energies
resulting from the left and right lead, respectively,
\begin{eqnarray}
\Sigma_{L,\ell}(\omega )=J_{L}^{2}\frac{2}{N}\sum_{k}\sin^{2}(k)
g_{k}(\omega
),\nonumber\\
\Sigma_{L,r}(\omega )=J_{R}^{2}\frac{2}{N}\sum_{p}\sin^{2}(p)
g_{p}(\omega ),
\end{eqnarray}
where $g_{k}(\omega ) $ and $g_{p}(\omega )$ are the Green
functions of the left and right leads, respectively. The two leads
are identical, except for being  connected to electronic
reservoirs with different chemical potentials, $\mu_{\ell}$ and
$\mu_{r}$. Hence,
\begin{eqnarray}
\Sigma_{L,\ell}^{<}(\omega )&=&f_{\ell}(\omega
)(\Sigma_{L,\ell}^{A}(\omega )-\Sigma_{L,\ell}^{R}(\omega )), \nonumber\\
\Sigma_{L,r}^{<}(\omega )&=&f_{r}(\omega )(\Sigma_{L,r}^{A}(\omega
)-\Sigma_{L,r}^{R}(\omega )),
\end{eqnarray}
with
\begin{eqnarray}
f_{\ell}(\omega )= \frac{1}{e^{\beta (\omega -\mu_{\ell})}+1},\ \
f_{r}(\omega )=\frac{1}{e^{\beta (\omega -\mu_{r})}+1}.
\end{eqnarray}
The superscripts $R$ and $A$ refer to the retarded and advanced
Green functions, respectively.

Our aim is to express the current through the interferometer in
terms of the dot Green function, $G_{dd}$, and the Green functions
of the ABI when the dot is disconnected, $g$. To this end, we will
use the Dyson equation
\begin{eqnarray}
G_{ij}=g_{ii}+X_{i}(\alpha )G_{dd}\widetilde{X}_{j}(-\alpha
),\label{DYSON}
\end{eqnarray}
where
\begin{eqnarray}
X_{i}(\alpha )&=&j_{\ell}e^{-i\alpha /2}g_{i-1}+j_{r}e^{i\alpha
/2}g_{i1},\nonumber\\
\widetilde{X}_{i}(-\alpha )&=&j_{\ell}e^{i\alpha /2}g_{-1
i}+j_{r}e^{-i\alpha /2}g_{1i}.
\end{eqnarray}
(Note that while the retarded and advanced Green functions $g$ are
symmetric in the site indices, this is not the case for the
Keldysh lesser functions, $g_{ij}^{<}$.) For brevity, we omit here
and below the $\omega$ dependence of the various functions.

Inserting Eq. (\ref{DYSON}) into Eqs. (\ref{IlIr}) for the
left-coming and right-coming currents, we obtain the currents
$I_{\ell}$ and $I_{r}$ in terms of $G^{<}_{dd}$ and $g_{ij}^{<}$.
The latter is given by
\begin{eqnarray}
g^{<}_{ij}=g^{R}_{i-n_{\ell}}
\Sigma^{<}_{L,\ell}g^{A}_{-n_{\ell}j}
+g^{R}_{in_{r}}\Sigma_{L,r}^{<}g^{A}_{n_{r}j}.
\end{eqnarray}
Then, using the relation
\begin{eqnarray}
g^{R}_{ij}-g^{A}_{ij}=[g^{R}(\Sigma^{R}_{L}-\Sigma^{A}_{L})g^{A}]_{ij},
\end{eqnarray}
\begin{widetext}
\noindent we obtain
\begin{eqnarray}
&&I\equiv\frac{I_{\ell}-I_{r}}{2}\nonumber\\
&=&\frac{e}{h}\int
d\omega\Bigl \{
(\Sigma^{A}_{L,\ell}-\Sigma^{R}_{L,\ell})\frac{1}{2} \Bigl
(G^{<}_{dd}+f_{\ell}(G^{R}_{dd}-G^{A}_{dd})\Bigr
)|X_{-n_{\ell}}^{R}(\alpha )|^{2}-
(\Sigma^{A}_{L,r}-\Sigma^{R}_{L,r})\frac{1}{2} \Bigl
(G^{<}_{dd}+f_{r}(G^{R}_{dd}-G^{A}_{dd})\Bigr
)|X_{n_{r}}^{R}(\alpha )|^{2}\nonumber\\
&+&(\Sigma^{A}_{L,r}-\Sigma^{R}_{L,r})
(\Sigma^{A}_{L,\ell}-\Sigma^{R}_{L,\ell})(f_{r}-f_{\ell}) \Bigl [
|g^{R}_{-n_{\ell
}n_{r}}|^{2}+G_{dd}^{R}g^{A}_{n_{r}-n_{\ell}}Z^{R}_{-n_{\ell}n_{r}}
+G_{dd}^{A}g^{R}_{n_{r}-n_{\ell}}Z^{A}_{-n_{\ell}n_{r}}\Bigr
]\Bigr \},\label{I}
\end{eqnarray}
where we have defined
\begin{eqnarray}
Z^{R}_{-n_{\ell}n_{r}}=j_{\ell}^{2}g^{R}_{n_{r}-1}g^{R}_{-n_{\ell}-1}
+j_{r}^{2}g^{R}_{n_{r}1}g^{R}_{-n_{\ell}1}
+j_{\ell}j_{r}\cos\alpha
(g^{R}_{n_{r}-1}g^{R}_{-n_{\ell}1}+g^{R}_{n_{r}1}g^{R}_{-n_{\ell}-1}).
\end{eqnarray}

The result (\ref{I}) holds in the general interacting case. When
there are no electronic interactions on the dot (as also
effectively happens in the slave-boson mean-field approximation),
this result can be simplified extensively. In that case,
$G_{dd}^{<}$ is known,
\begin{eqnarray}
G^{0<}_{dd}=G^{0R}_{dd}\Sigma^{0<}_{dd}G^{0A}_{dd},
\end{eqnarray}
with
\begin{eqnarray}
\Sigma_{dd}^{0<}&=&f_{r}|X_{n_{r}}^{R}(\alpha
)|^{2}(\Sigma_{L,r}^{A}-\Sigma_{L,r}^{R})+f_{\ell}|X^{R}_{-n_{\ell}}(\alpha
)|^{2}(\Sigma^{A}_{L,\ell}-\Sigma^{R}_{L,\ell}).
\end{eqnarray}
In addition, when there are no interactions present,
\begin{eqnarray}
G^{0R}_{dd}-G^{0A}_{dd}=G^{0R}_{dd}G^{0A}_{dd}(\Sigma^{0R}_{dd}-\Sigma^{0A}_{dd}),
\end{eqnarray}
with
\begin{eqnarray}
\Sigma^{0R}_{dd}-\Sigma^{0A}_{dd}=|X_{-n_{\ell}}(\alpha
)|^{2}(\Sigma^{R}_{L,\ell}-\Sigma^{A}_{L,\ell})+ |X_{n_{r}}(\alpha
)|^{2}(\Sigma^{R}_{L,r}-\Sigma^{A}_{L,r}).
\end{eqnarray}
Using these expressions in Eq. (\ref{I}), we obtain the current
through the non-interacting ABI, $I^{0}$,
\begin{eqnarray}
I^{0}&=&\frac{e}{h}\int d\omega
(\Sigma^{A}_{L,r}-\Sigma^{R}_{L,r})
(\Sigma^{A}_{L,\ell}-\Sigma^{R}_{L,\ell})(f_{r}-f_{\ell}) \Bigl [
|g^{R}_{-n_{\ell
}n_{r}}|^{2}+|G^{0R}_{dd}|^{2}|X^{R}_{-n_{\ell}}(\alpha
)^{2}|X^{R}_{n_{r}}(\alpha )|^{2}\nonumber\\
&+&G^{0R}_{dd}g^{A}_{n_{r}-n_{\ell}}Z^{R}_{-n_{\ell}n_{r}}
+G^{0A}_{dd}g^{R}_{n_{r}-n_{\ell}}Z^{A}_{-n_{\ell}n_{r}}\Bigr
]\Bigr \}\nonumber\\
&=&\frac{e}{h}\int d\omega (\Sigma^{A}_{L,r}-\Sigma^{R}_{L,r})
(\Sigma^{A}_{L,\ell}-
\Sigma^{R}_{L,\ell})(f_{r}-f_{\ell})|G^{0R}_{-n_{\ell}n_{r}}|^{2},\label{SOF}
\end{eqnarray}
where in the last step we have used Eq. (\ref{DYSON}). In our
tight-binding description of the semi-infinite identical leads,
\begin{eqnarray}
\Sigma_{L,\ell (r)}^{A}-\Sigma_{L,\ell (r)}^{R}=i\frac{2J^{2}_{L
(R)}}{J}\sqrt{1-\Bigl (\frac{\omega}{2J}\Bigr )^{2}}.
\end{eqnarray}
Using this in Eq. (\ref{SOF}), together with
$f_{r}-f_{\ell}=(\mu_{r}-\mu_{\ell})(\partial f /\partial \omega)$
leads to the Landauer formula, Eq. (\ref{LANDAUER}).

\end{widetext}



\begin{thebibliography}{999}
\bibitem{note1} The non-interacting  self-energy of the
dot  (which determines the Kondo temperature) in this special case
is $\Sigma_{dd}^0(\omega)=2j_l^2 [\cos^2(\al/2)
g_1^e(\omega)+\sin^2(\al/2) g_1^o(\omega)]$, where $g_1^{e/o}$ are
the non-interacting Green functions at the extremity of the
even/odd lead. In the very long wire limit, $n_l\gg \xi_K^0\gg 1$,
one may use a continuum approximation in which the very fast
variations of $g_1^{e/o}(\omega)$ are replaced by their (energy)
average values, $\rho_1^e=-Im(g_1^e)/\pi \approx const. $ and
$\rho_1^o=-Im(g_1^o)/\pi\approx const.$,  and the real part of
$\Sigma_{dd}^{0}$ is neglected. Then, $\Sigma_{dd}^0$ becomes
phase-independent.
\bibitem{note2} Note that the variations of $T_K^{SBMFT}$
with the flux, in the case of a large-size ABI,  are  mainly due
to the flux-dependence of $\Delta$ (the imaginary part of the
non-interacting self-energy). The  slave boson parameter $b_0$ is
almost constant for the $L=403$ ABI, as opposed to the $L=21$ one.
\end{thebibliography}
\end{document}